# Monostable Super Antiwettability

Yanshen Li[1,2], Cunjing Lv[1,3], David Quéré[4*], Quanshui Zheng[1,2,5*]

[1]Department of Engineering Mechanics, Tsinghua University, Beijing 100084, China.

[2]Center for Nano and Micro Mechanics, Tsinghua University, Beijing 100084, China.

[3]Institute for Nano- and Microfluidics, Center of Smart Interfaces, Technische Universität Darmstadt, Alarich-Weiss-Straße 10, 64287 Darmstadt, Germany.

[4]Physique et Mécanique des Milieux Hétérogènes, UMR 7636 du CNRS, ESPCI, 75005 Paris, France.

[5]Applied Mechanics Lab, and State Key Laboratory of Tribology, Tsinghua University, Beijing 100084, China.

*E-mail: zhengqs@tsinghua.edu.cn, david.quere@espci.fr

**Super-antiwettability is an extreme situation of wetting where liquids stay at the tops of rough surfaces, in the so-called Cassie state[1]. Owing to the dramatic reduction of solid/liquid contact, it has many applications, such as antifouling[2,3], droplet manipulation[4,5], and self-cleaning[6-9]. However, super-antiwettability is often destroyed by impalement transitions caused by environmental disturbances[10-16] while inverse transitions without energy input have never been observed[12,17-21]. Here we show through controlled experiments that there is a "monostable" region in the phase space of the receding contact angle and roughness parameters where transitions between (impaled) Wenzel and Cassie states can be reversible. We describe the transition mechanism and establish a simple criterion that predicts the experimentally observed Wenzel-to-Cassie transitions for different liquids placed on micropost-patterned substrates. These results can guide for designing and engineering robust super-antiwetting surfaces.**

Wetting on hydrophobic textured substrates is usually bistable[10,20]: two states can be observed, depending on the history of the system owing to the existence of an energy barrier between them. A drop gently deposited on the texture will be often in the Cassie levitating state, while the same drop can get impaled in the Wenzel state after an impact. Most generally, one of these two states is metastable[10,22], and the barrier between them is large enough to prevent spontaneous transitions[22,23]. Metastability is beneficial for achieving relatively robust slippery Cassie states, but detrimental for strongly anchored sticky Wenzel states[10]. In this context, it is worth exhibiting "monostable" Cassie state where even an accidental transition to the Wenzel state (due to force fluctuations such as encountered in an impact, or to pressure applied on the liquid) can be repaired by the absence of barrier between both states. Since reported Wenzel-to-Cassie (W2C) transitions generally involve either an external energy input[12,17-19,21], or a potential energy release[20,24-26], such monostable Cassie states can be seen as unreal.

In this paper, we test the existence and robustness of Cassie and Wenzel states by pressing a water drop of radius $R$ against a textured substrate made of silicon and etched so as to be decorated by a square array of square microposts with side $a = 19\ \mu$m, height $h = 100\ \mu$m and spacing $b = 101\ \mu$m (Figure 1). At these scales and post densities, Cassie and Wenzel states can be distinguished using backlighting and side views: a thin ray of light below the liquid is the signature of the Cassie state. In addition, we place under the textured material a force captor (Mettler Toledo, XA205, Switzerland), which allows us to monitor the force exerted by the drop as it is squeezed against its substrate.

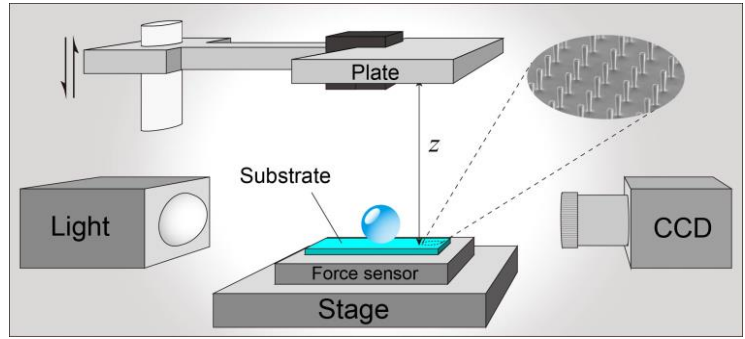

**Figure 1.** Schematic of the experimental setup. A millimeter-size water drop is placed on the hydrophobic textured material shown in the insert (pillars with side $a = 19$ μm, height $h = 100$ μm and spacing $b = 101$ μm). The drop is pressed by a hydrophobic plate (150 mm x 30 mm x 5 mm), which induces a transition from the Cassie state to the Wenzel state. The plate is moved down a little bit further, before reversing its motion. Side views allow us to extract the drop profile and surface energy as a function of the distance $z$. Simultaneously, the force exerted on the substrate is measured with a precision balance (force sensor).

We compare in Figure 2 the usual case of an irreversible C2W (Cassie-to-Wenzel) transition (Fig. 2a-b) with a situation of monostable Cassie state (Fig. 2c-d). In a first series of experiments, the substrate is treated by OTS. The corresponding advancing and receding contact angles of water on such surfaces yet flat are respectively $\theta_a = 113 \pm 3°$ and $\theta_r = 91 \pm 2°$. A water drop (of radius $R = 1$ mm, smaller than 2.7 mm, the capillary length of water) is placed on the substrate in the Cassie state and then slowly ($dz/dt = 50$ μm/s, in absolute value) pressed against it using an OTS-treated glass plate (Fig. 2a). As the distance $z$ between both plates quasi-statically decreases, the drop undergoes a C2W transition, as also seen in Fig. 2b where the Wenzel radius $R_W$ increases abruptly from 0 to 1.3 mm at $z = 430$ μm. The drop impales where it sits, so that $R_S$, the apparent radius of contact with the substrate is continuous at the transition. After further pressing (down to $z = 310$ μm), the movement of the upper plate is reversed at a lower constant speed $dz/dt = 20$ μm/s (red data in Fig. 2b). However, the drop remains irreversibly pinned ($R_S = R_W$), so that its radius modestly decreases as plates are separated.

The sequence of events is markedly different if the receding angle of the liquid on the textured solid is significantly higher, that is, with weaker pinning. This situation was obtained on the same substrate as in Fig. 2a-b either with mercury, or with water after adding to the substrate a fine nanotexture made of colloids with radius of 30 nm (Glaco treatment). These two cases provide respective advancing and receding contact angles $\theta_a = 165 \pm 2°$ and $\theta_r = 137 \pm 4°$, $\theta_a = 164 \pm 2°$ and $\theta_r = 150 \pm 3°$. Results obtained in these two situations are similar, and we report in Fig. 2c-d the behavior a water drop pressed and released against textures with a Glaco subtexture. The C2W transition to the Wenzel state in the pillars is observed for spacing $z$ smaller than that in the first experiment, and differences are mainly found in the releasing stage: on the one hand, the curve followed for the contact radius $R_S$ (triangular symbols in Fig. 2d) is reversible; on the other hand, the Wenzel radius $R_W$ continuously decreases in a narrow interval of separation distances $z$, so that a full Cassie state is recovered for a distance $z$ as small as 350 µm.

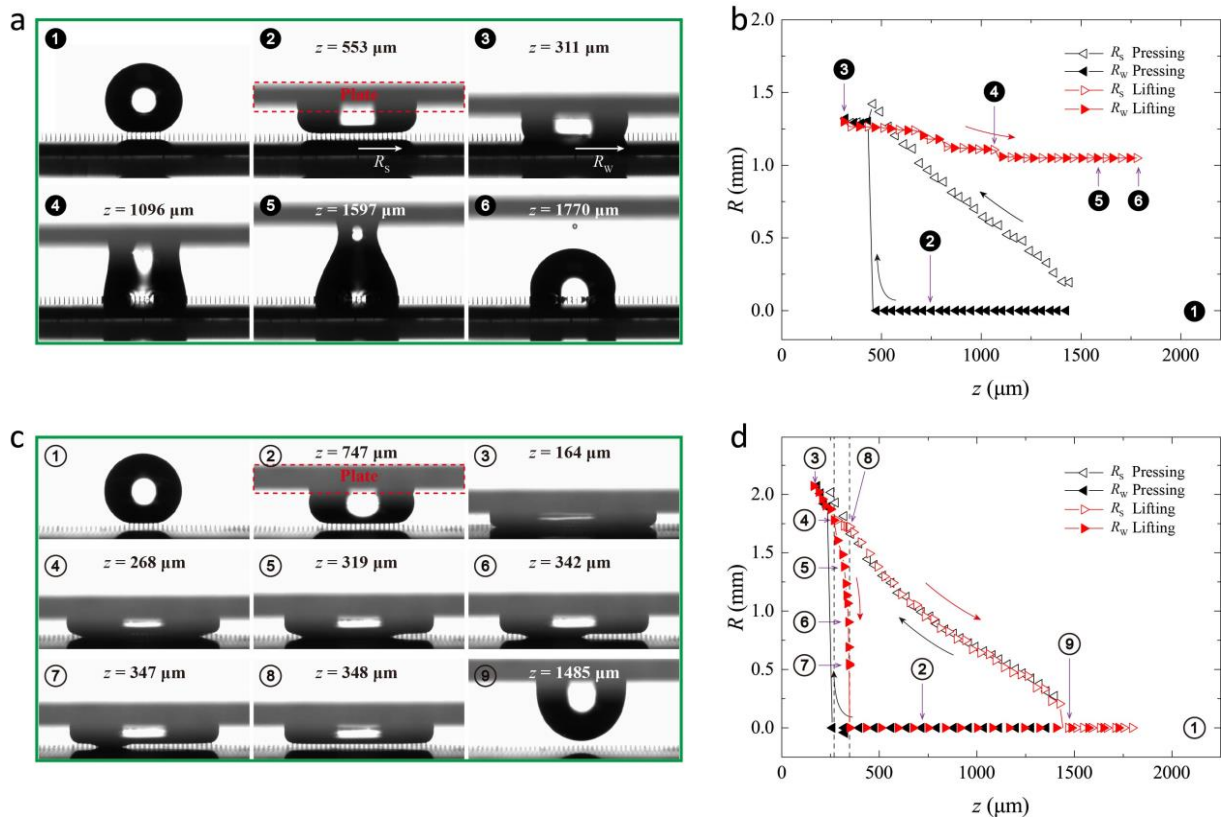

**Figure 2.** States and contact radii of a drop squeezed against a textured surface. **a.** Six snapshots for an experiment performed with water (with $R = 1$ mm) on silicon pillars with height and spacing of 101 µm treated with OTS. Silicon treated with OTS yields $\theta_a = 113 \pm 3°$ and $\theta_r = 91 \pm 2°$. Corresponding movie is Supplementary Movie 1. The drop initially in the Cassie state (❶) is pressed (❷) until it reaches the Wenzel state (❸). Upon lifting, the drop gets stretched (❹,❺) yet strongly pinned in the Wenzel state (❻). **b.** Drop radius as a function of the distance $z$ between the plate and substrate. $R_s$ (with empty triangles) denotes the radius of contact between the drop and the top of the substrate, and $R_w$ (with full trangles) denotes the radius of the Wenzel contact. Symbols are black and left pointed in the pressing stage, and they are red and right pointed in the lifting stage. The states corresponding to Fig. 1a are marked with numbers. **c.** Same experiment after treating the pillars of Fig. 2a with Glaco nanotextured coating. Water contact angles then become $\theta_a = 164 \pm 2°$ and $\theta_r = 150 \pm 3°$. Corresponding movie is Supplementary Movie 2. The initial Cassie drop (①) is pressed (②) so to reach the Wenzel state (③). Upon lifting, the drop first shrinks (④), then water dewets the pillars gradually from the contact line (⑤⑥⑦), forming a mixed state that eventually fully dewets the pillars; the corresponding compressed Cassie state (⑧) remains unchanged over further lifting (⑨). **d.** Drop radii $R_w$ and $R_s$ as functions of the squeezing distance $z$. The two vertical dashed lines indicate the boundaries of the mixed state and numbers refer to Fig. 2c.

As drops are pressed and released, they exert a force denoted as $F$ on the substrate. $F$ can be measured with a force sensor, as sketched in Fig. 1. Data acquisition frequency is 2Hz. In addition, we can extract from the succession of photos (such as displayed in Fig. 2) the surface areas of the different interfaces, and then calculate for each $z$ the free energy (the total of surface and interfacial energies), $U$, of the drop. Solid-liquid interface tensions are deduced from Young's relationship: $\gamma_{SL} - \gamma_{SV} = -\gamma\cos\theta_E$, where we chose for the equilibrium angle $\theta_E$ the mean (measured) contact angle of water $(\theta_a + \theta_r)/2$ on the corresponding surface, that is, $\theta_E = 97°$ on OTS treated glass plate, $\theta_E = 102°$ on OTS treated silicon, $\theta_E = 97°$ on OTS treated glass plate and $\theta_E = 157°$ on Glaco treated silicon substrate. We plot in Figure 3 the variations of both $F$ and $U$ as a function of $z$, for the two experiments reported in Fig. 2.

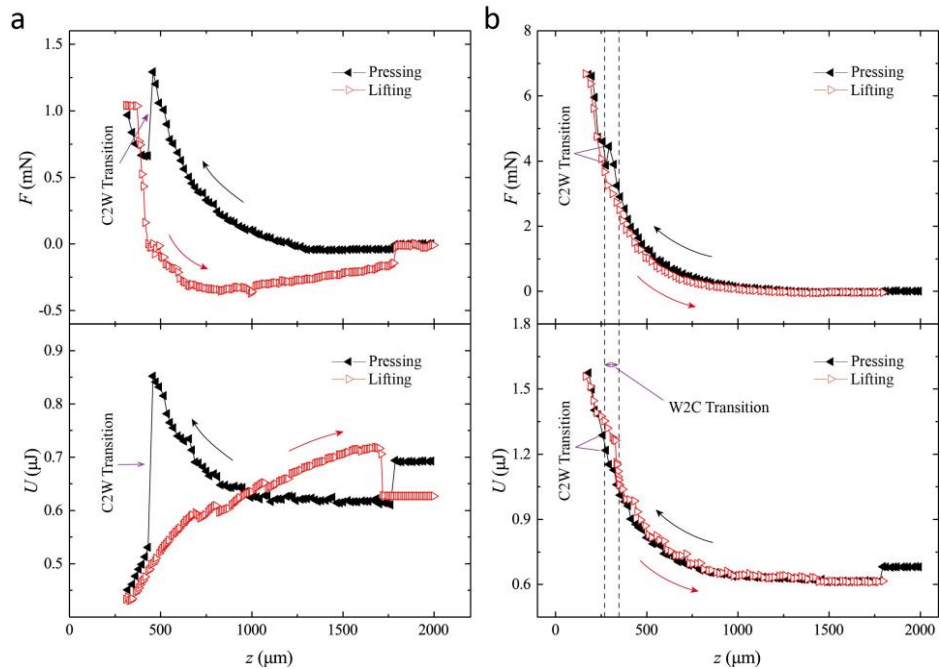

**Figure 3.** Force and energy variation in the two experiments reported in Figure 2. The force $F$ exerted by the drop on its substrate is measured with a force captor, and plotted as a function of the squeezing distance $z$; the interfacial energy $U$ of the deformed drop is deduced from image analysis, and also plotted as a function of $z$. Data are full and black in the pressing stage, and empty and red in the releasing stage. **a**. Force and energy on the OTS-treated surfaces, corresponding to the experiments in Fig. 2a-b. **b.** Force, pressure and energy on the Glaco-treated surfaces, corresponding to the experiments in Fig. 2c-d. The vertical dashed lines mark the boundaries of the mixed state of the drop. The pressure is deduced from the measured principal radii of curvature of the drop, after using the Laplace formula.

Figure 3 confirms and makes more quantitative the observations in Figure 2. Instead of a (classical) irreversible transition C2W with marked hysteresis loop (Fig. 3a), we observe on low adhesive substrates quasi-reversible trajectories (Fig. 3b). In both cases, the pressing stages are first quite similar: the force increases as the Cassie drop is squeezed, corresponding to flatter and flatter shapes. The bare surface energy $U_0$ of a water drop of radius $R = 1$ mm is $4\pi R^2 \gamma$, that is, 0.7 μJ, and the interfacial energy of the drop as it is squeezed typically increases by a comparable magnitude ($\Delta U \approx 0.3$ μJ). The plots however markedly differ at the C2W transition: in case a, the force and energy both drop in a discontinuous fashion, which highlights the existence of a large energy barrier ($\sim U_0$) between the two states. Then the drop is strongly pinned in the texture, which can be seen from the negative values of the force in the releasing stage: stretched water resists the force tending to extract it out of the texture. This capacity of water to resist negative pressure is also reflected by a continuous rise of energy as $z$ increases in the releasing stage. All these behaviors are in sharp contrast with what can be observed on a substrate with less pinning (Figure 3b): on the one hand, both curves $F(z)$ and $U(z)$ are continuous at the W2C and C2W transitions; on the other hand, pressing and releasing stages are nearly superimposed. In particular, $F(z)$ remains positive all along the trajectory. Finally, the energy landscape shows that the total energy of the drop similarly decreases monotonously during the lifting process, while the droplet undergoes a transition from the pressed Wenzel state to Cassie state. This energy landscape shows that the Cassie state is monostable, rather than metastable on classical superhydrophobic surfaces.

As the W2C transition zone expands (or equivalently, as the radius of the locally Wenzel zone decreases from $R_W$ to $R_W - \Delta R_W$), the free energy changes (see Supplementary Section 5 for

a detailed derivation) as follows:

$$\Delta U = [1 - f + (r - f)\cos\theta_r]g(2\rho R_W \Delta R_W)$$

where $f$ and $r$ are the areas of the pillars' tops and the rough substrate per unit apparent substrate area, respectively; and $\theta_r$ is is the receding contact angle of the substrate material. Because the wetted area during the lifting process is receding, the receding contact angle, $\phi_r$, rather than the Young's contact angle acts, and the energy dissipation in moving the three-phase contact lines has been considered by $\theta_r$. A sufficient condition for realizing automatic W2C transition is $\Delta U < 0$ throughout the transition process. Therefore, from the above relation, we obtain the following criterion of the monostable Cassie state:

$$\frac{1-f}{r-f} < -\cos\theta_r. \qquad (1)$$

Interestingly, the criterion for a globally stable Cassie state (i.e., a lower free energy than in the Wenzel state) has a very similar form ($\frac{1-f}{r-f} < -\cos\theta_E$ to that of Eq. (1)[#7]. Because the Young's contact angle $\theta_E$ is always larger than the receding angle $\theta_r$, the above two relationships are consistent with an obvious requirement that the global stability is a necessary condition for monostability.

We note that the above-mentioned monostable and non-monostable Cassie states for water agree with the criterion (1). To validate criterion (1) in a wider range, we conduct a series of similar experiments to those shown in Figs. 1 and 2 with varying $\theta_r$, $f$ and $r$. In these experiments, silicon substrates patterned by square array of square or circular microposts with a fixed side length or diameter (i.e., $a = 100$ μm and height $h = 100$ μm, respectively) are used, and the pitch $b$ is varied to change $f = a^2/(a+b)^2$ or $f = \pi a^2/4(a+b)^2$ (see Supplementary Section 6). To have a wider choice of $\theta_r$, mercury is used as the testing liquid instead of water. Different surface treatments of the substrates are used to change the contact angles of the mercury droplets (see Methods section), and copper plates with nanostructured CuO on the surface are used as the pressing plate, since they have no contact angle hysteresis with mercury (see Supplementary Section 7). The open and solid dots in the square (i.e., square pillars) and circular (i.e., circular pillars) shapes in Fig. 3a show the results are experimentally observed to be monostable and non-monostable, respectively. In comparison, the green area above the dashed line in Fig. 3a corresponds to criterion (1), indicating the monostable Cassie region. Thus, criterion (1) agrees well with the experimental observations. As a comparison, region above the red dash-dotted lines in Fig. 3 correspond to the global stable criterion of Cassie state. This criterion is clearly a necessary, but not a sufficient condition for ensuring monostable Cassie state.

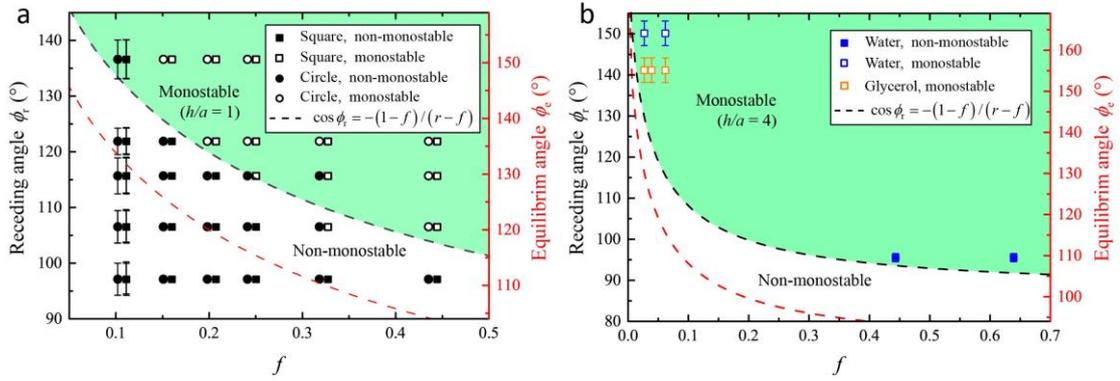

**Figure 4.** Experimentally observed monostable Cassie (open dots) and bistable Cassie (solid dots) cases in the ($\phi_r, f$) phase space on micropost-patterned substrates fabricated from silicon wafers. **a**, The microposts are square-shaped with a fixed height $h$=100 μm and a side length $a$=100μm. The pitch (the post separation) $b$ is varied to provide different area fractions $f = a^2/(a+b)^2$ and roughnesses $r = 1 + \frac{h}{s}f$. Each row consists of six data points, which correspond to six small substrates taken from one large ion-etched silicon wafer and treated concurrently to ensure the same surface properties including contact angle. Receding contact angles are averaged over six measurements, and the error bars (the same for other area frictions $f$) indicate the standard deviations of the data. **b**, Experiments with water and glycerol. The microposts are square-shaped with a fixed height $h$=100 μm, a side length $a$=19 μm, and different separations $b$. Receding contact angles are averaged over five measurements. Error bars indicate the standard deviations of the data.

Criterion (1) should be generally valid for various liquids and rugged substrates. To find supporting evidence, we fixed $a = 19\ \mu$m and $h = 100\ \mu$m, and varied $b$ for the micropost-patterned substrates so that the area fractions $f$ are similar to those used in Fig. 3a, but the ruggedness $r = 1 + \frac{h}{s}f$, or equivalently the height/side length ratio $\frac{h}{a}$, is different. Figure 3b shows the experimental results with open and solid dots indicating monostable and non-monostable Cassie states for water (blue dots) and glycerol (orange dots, see Supplementary Section 8). The green area above the dashed line in Fig. 3b is the monostable Cassie region predicted by criterion (1). Again, we see excellent agreement between the theoretical and experimental results. Unlike mercury, it is challenging to get more contact angles with water and glycerol to meet criterion (1).

An additional interesting observation from the insert of Fig. 2d is that the Laplace pressures, $p_{W2C}$, are nearly constant throughout the W2C transition stage. To understand, we first note that there is another critical Laplace pressure, $\Delta p_{C2W}$, which is the maximum pressure before the pillars will pierce into the liquid (or equivalently the Cassie state will transform to the Wenzel state). This pressure can be precisely formulated as[27]:

$$\Delta p_{C2W} = -\frac{f}{1-f}\frac{\gamma}{s}\cos\phi_a \qquad (2)$$

where $\phi_a$ is the advancing contact angle of the substrate's surface, and $s$ is the ratio of the area to the perimeter of the pillars' cross-sections. This result is valid for any micropost's cross-sectional shape. For both square- and circular-shaped cross-sections with a side length or diameter $a$, $s = a/4$. The observation $\Delta p_{W2C} < \Delta p_{C2W}$ is consistent with the physical concepts of $\Delta p_{W2C}$ and $\Delta p_{C2W}$. After a small expansion of the W2C transition zone, a small amount of liquid is drained from under the microposts' tops to above the droplet, which increases the Laplace pressure until the plate is further lifted. In this interval, the increased Laplace pressure will stop the W2C from expanding but remains below the critical pressure $\Delta p_{C2W}$ so that the W2C transition zone can persist. After further upward movement of the plate, the Laplace pressure decreases to the value $p_{W2C}$, which leads to the W2C transition expansion. This logical process is experimentally confirmed: when we stop the lifting in the W2C transition stage, we observed the stop of the transition (see Supplementary Section 9). We therefore call $\Delta p_{W2C}$ the W2C transition pressure. In the W2C transiting zone, the process of the liquid surface is inverse to that in the piercing one and is thus a receding motion. Thus, using a similar derivation (see Ref. 27), we can obtain:

$$\Delta p_{W2C} = -\frac{f}{1-f}\frac{\gamma}{s}\cos\phi_r \qquad (3)$$

This prediction, which is described by the red dashed line in Fig. 2d, agrees well with the experimental results.

To conclude, we first note that the above results do not consider the effect of droplet size. We find experimentally that $\Delta p_{W2C}$ is independent of the liquid droplet volume in most cases. However, for a small droplet with a radius $R_{sph}$, as the drop of the same volume is spherical, the Laplace pressure $p_{sph} = \frac{2\gamma}{R_{sph}}$ can exceed the W2C transition pressure (3). Because the droplet is nearly spherical in the Cassie state, droplets must have larger radii than the following critical radius to achieve a monostable Cassie state:

$$R_{sph,cr} = -\frac{1-f}{f}\frac{a}{2\cos f_r}. \qquad (4)$$

This property is independent of the liquid and micropost heights used. For example, when $f = 0.1$ and $\phi_r = 130°$ or $150°$, we obtain from equation (4) that $R_{sph,cr} = 7.0a$ or $5.2a$, respectively. Thus, smaller $a$ and $f$ and larger $\phi_r$ allow smaller droplets to achieve monostable Cassie contact. The above result is particularly inspiring for considering super antiwettability in condensation or fogging because in a condensation process water droplets are typically nucleated at tens of nanometers in size[29]. Therefore, to achieve monostable Cassie state in condensation and fogging, we may have to use nanoposts, instead of microposts. This is the practice of many naturally occurring superhydrophobic surfaces such as lotus leave.

Finally, by noting that the receding angles of water on flat surfaces are rare to exceed 120°[30],

we are not sure at this moment whether or not there would be substrates that could realize monostable Cassie state with only one-level ruggedness. Nevertheless, as we showed before we can achieve the monostable Cassie state for water by using two-level ruggedness as presented in most of naturally occurring superhydrophobic surfaces.